\documentclass[aps,prl,twocolumn,groupedaddress]{revtex4}
\usepackage{graphicx}
\usepackage{amssymb}
\begin{document}

%\preprint{}

\title{Information hiding and retrieval in Rydberg wave packets using half-cycle pulses}


\author{J. M. Murray$^1$, S. N. Pisharody$^1$, H. Wen$^1$, C. Rangan$^2$, P. H. Bucksbaum$^1$}
%\thanks{}
\affiliation{$^1$ Department of Physics, University of Michigan, Ann Arbor, MI 48109-1120}
\affiliation{$^2$ Department of Physics, University of
    Windsor, ON N9B 3P4, Canada.}

%\altaffiliation{Department of Physics,
%University of Windsor, ON N9B 3P4}


\date{\today}

\begin{abstract}

We demonstrate an information hiding and retrieval scheme with the relative phases between states
in a Rydberg wave packet acting as the bits of a data register. We use a terahertz half-cycle
pulse (HCP) to transfer phase-encoded information from an optically accessible angular momentum
manifold to another manifold which is not directly accessed by our laser pulses, effectively
hiding the information from our optical interferometric measurement techniques. A subsequent HCP
acting on these wave packets reintroduces the information back into the optically accessible data
register manifold which can then be `read' out.

\end{abstract}

%\keywords{}

\maketitle


Coherent excited states of multilevel quantum systems have been proposed as quantum bit registers
for storing and manipulating information \cite{Levine,StenholmPRA01,shapiroPRL03}. Previous work
has demonstrated the use of the quantum phases of the states in a Rydberg wave packet for
information storage and showed the applicability of terahertz HCP's to retrieve and manipulate
this information \cite{ahn,ahn2,ahn3,rangan:033417,RanganModOpt2002,RanganPRA2005}. Recently, we
have measured the effects of an HCP on the phases and amplitudes of Rydberg $p$-state data
registers\cite{murrayPRA05}. In the current work, we extend this investigation to the possibility
of performing sequential coherent operations on the stored information by using multiple HCP's.
Since any program is a sequence of operations, the ability to perform multiple operations on all
or selected parts of the stored data is an essential requirement for information processing.

We use a pair of HCP's acting on a Rydberg wave packet to demonstrate the storage and retrieval of
information from the wave packet. In our experiments, information is stored in the phases of each
of the states of the wave packet, with respect to the phase of a reference state. The first HCP
acts on the wave packet storing the information, redistributing the complex probability amplitudes
of the states in a deterministic manner\cite{murrayPRA05}. Our means of detecting the stored
information is a wave packet holography technique that has previously been applied to wave packet
sculpting\cite{weinachtNature99}. This technique uses interference with an $\ell=1$ reference
Rydberg wave packet, so it is sensitive only to the $p$-state populations in the wave packet. The
non-$p$ states populated by the HCP are therefore hidden from measurement. A subsequent HCP can
redistribute the $\ell\neq 1$ state populations back into the $p$-states and make the information
available for measurement. This ability to hide selective parts of the information and retrieve it
at will allows us to introduce operators timed to occur between the two HCP's enabling us to act
on a subset of the stored information. Here, we report the demonstration of this information
hiding and retrieval scheme.

A tightly focused $1079\,\mathrm{nm}$ pulse from a Ti:Sapphire-pumped optical parametric amplifier
excites ground state cesium atoms from an effusive source from the $6s$ state into an intermediate
$7s$ launch state. A spectrally shaped $800\,\mathrm{nm}$ pulse excites an $n=27,\dots,32$
$p$-state Rydberg wave packet with equal phases and approximately equal amplitudes\cite{ahn3}.
This wave packet is subsequently kicked by a weak THz HCP polarized along the same direction as
the laser pulses. The duration of the HCP (0.5ps) is significantly shorter than the Kepler period
($\sim$ 3ps). This suggests that an impulse approximation can be used, and theory and experiment
confirm this\cite{ahn2}. The HCP impulsively transfers a momentum $Q = 0.0017\,\mathrm{a.u.}$
(atomic units) to the Rydberg electron. This momentum kick transfers population from an initial
$\ell = 1$ manifold into other $p$-states as well as $\ell\neq 1$ angular momentum states. With
the correct choice of HCP delay $T_1$, we can selectively depopulate one of the
states\cite{murrayPRA05}.
%and, in so doing, move the information contained in that phase into
%states of other angular momentum\cite{murrayPRA05}. To retrieve that
%information,
A second HCP is applied at time $T_2$, transferring amplitude from non-$p$ states back into the
$\ell = 1$ manifold. The resultant changes in phase for those $p$-states are then measured by
exciting the reference $p$-state Rydberg wave packet at different delays, $\tau$, with a second
ultrafast pulse, identical to the first. The state selective field ionization (SSFI) spectrum is
used to analyze the interference between the wave packets and determine the phase relationships
between the states in the wave packet. Details of this phase measurement procedure have been
described previously\cite{murrayPRA05}.

Information stored in the phases of the states of the wave packet is retrieved through correlation
measurements. For two $p$-states $\left|j\right>$ and $\left|k\right>$, the noise-free correlation
between their populations is given by
\begin{eqnarray}\label{correlation}
    r_{jk}(\tau) & = &
    \cos((\phi_{j1}-\phi_{k1})-(\phi_{j2}-\phi_{k2})-(\omega_j-\omega_k)\tau)\nonumber\\
    & = & \cos(\Phi_{jk}-\omega_{jk}\tau).
\end{eqnarray}
Here, $\phi_{j1}$ is the phase with which the state $\left|j\right>$ is excited, a time $\tau$
before the reference pulse excites the $j$-component of the reference wave packet with phase
$\phi_{j2}$. The Rydberg state frequency for state $\left|j\right>$ is denoted by $\omega_j$. The
amplitude of this correlation curve is unity in the absence of technical noise and decoherence.
The presence of decoherence and background noise result in a measured correlation
amplitude\cite{murrayPRA05}
\begin{equation}\label{correlation2}
    r_{jk}^{meas}=\sqrt{\left(1-\frac{\sigma_{N_j}^2}{\sigma_{j_{meas}}^2}\right)
    \left(1-\frac{\sigma_{N_k}^2}{\sigma_{k_{meas}}^2}\right)}
     \cdot r_{jk},
\end{equation}
where $\sigma_{N_j}$ is the standard deviation of the noise present in the measurement of state
$\left|j\right>$ and $\sigma_{j_{meas}}$ is the standard deviation of the population identified as
state $\left|j\right>$. Noise can also introduce an uncertainty $\Delta\Phi_{jk}$ in the phase of
$r_{jk}$.

The correlation amplitude $r_{jk}$ is a periodic function of the reference time delay, $\tau$
(Eq.\ref{correlation}). The measured correlation curve is described primarily by its phase and its
amplitude. The effect of a HCP is to modify the phases and amplitudes of the correlation curves.
The amplitudes and phase shifts of the correlations due to the double HCP kick depend on the
relative phases between the $p$-states, and hence on the HCP delays $T_1$ and $T_2$.

\begin{figure}[t]
\includegraphics[width=\columnwidth]{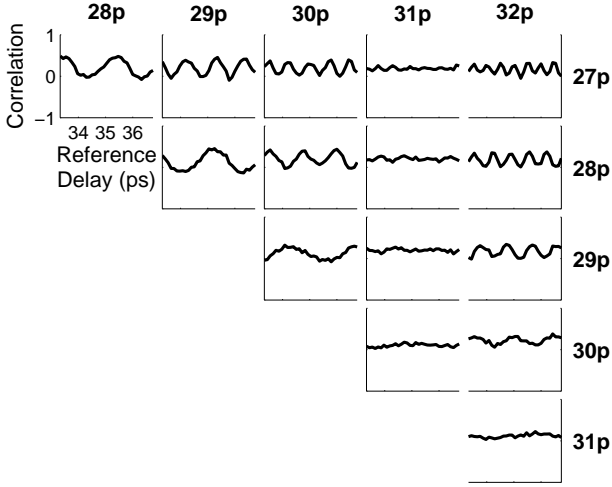}
\caption{\label{corr_vanish} Correlations vanish} The choice of
$HCP_1$ delay $T_1=5\,\mathrm{ps}$ moves population from the $31p$
state into neighboring angular momentum states such that the phase
information that was contained in $31p$ cannot be measured. All
correlations involving $31p$ vanish.
\end{figure}
\begin{figure}[t]
\includegraphics[width=\columnwidth]{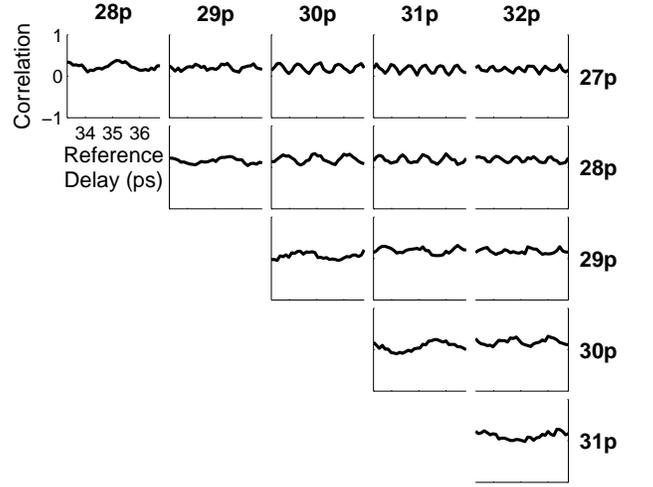}
\caption{\label{corr_recovered} Correlations recovered} Applying a
second HCP at $T_2=6.3\,\mathrm{ps}$ coherently transfers population
from neighboring angular momentum states into the data register
state, $n=31p$. The correlations with $31p$ are seen to return;
phase information is recovered.
\end{figure}

For information processing purposes, it is useful to quantify the amount of information that can
be reliably retrieved, both with and without an HCP kick. When storing information in the quantum
phase, $\phi_k$, of a state $\left|k\right>$ (with respect to some reference), one can divide the
phase range, $[0,\ 2\pi)$ into $N$ different partitions, each spanning a phase of $2\pi/N$ and
representing a different discrete logical level\cite{Preskill}. For example, if we have 10
partitions, the scheme would be a decimal system. The information capacity of any digital encoding
scheme with $N$ logical levels scales as $\log N$\cite{Brillouin,Frank,Lloyd}. The number of
logical levels distinguishable in the quantum phase is related to the precision with which the
phase can be measured, $\phi_k\pm\frac{\Delta\phi_k}{2}$. If the uncertainty in phase,
$\Delta\phi_k$, is zero, the phase (and the encoded information) is known exactly; if the
uncertainty is $2\pi$, nothing is known about the phase. The information capacity of a state with
phase uncertainty, $\Delta\phi_k$ can be written as
\begin{equation}
i_k = \log \frac{2\pi}{\Delta\phi_k}. \label{eqn:contin_info}
\end{equation}

The state phases, $\phi_k$ containing the encoded information are
extracted using the correlation technique. We measure the
correlation between all pairs of states.  The phase associated with
each state can be calculated as the difference in phases from
multiple correlation measurements (for example,
$\phi_k=\Phi_{k,\mathrm{ref}}$, and
$\phi_k=\Phi_{k,j}-\Phi_{j,\mathrm{ref}}$ for all $j$). The noise
level in these measurements influences the phase uncertainty
associated with each phase difference. The quantity $\Delta\phi_k$
is measured as a weighted average of the phase uncertainties in the
measured correlation curves,
\begin{eqnarray}
\Delta\phi_k = \frac{\sum_j W_{jk}\Delta\Phi_{jk}}{\sum_j W_{jk}}= \frac{N-1}{\sum_j
1/\Delta\Phi_{jk}},
\end{eqnarray}
where $W_{jk}=1/\Delta\Phi_{jk}$ is the weighting factor in each equation and the sum over the
weights is included for normalization. This phase uncertainty is used as described above (Eq.
\ref{eqn:contin_info}) to determine the amount of information retrieved from the measurement of
that phase. To obtain the fidelity of information retrieval across the entire wave packet, we add
the information capacity of each state so that the total information content is
\begin{equation}
I=\sum_k i_k=\sum_k \log \frac{2\pi}{\Delta\phi_k}.
\end{equation}

We perform our experiment with a Rydberg wave packet excited into $np$ states with $n=27\dots 32$.
A correlation measurement on this wave packet allowed us to determine $\Delta\phi=7^\circ$
($i=5.7$ bits) for the $31p$ state. Fig.\ref{corr_vanish} shows a correlation measurement when an
HCP is applied at $T_1=5\,\mathrm{ps}$ such that the correlation vanishes for all pairs of states
involving $31p$. At this particular delay, population in state $31p$ has been substantially
transferred to other states, both $\ell = 1$ and $\ell \neq 1$. Much of the phase information that
was initially stored in the $31p$ state is now inaccessible to our measurement technique, which is
only sensitive to the $p$-states. The phase uncertainty in the $31p$ state is $32^\circ$ ($i=3.5$
bits) from this measurement.

In the absence of a half-cycle pulse, the information capacity of the wave packet as determined
using the correlation technique (Eq.\ref{eqn:contin_info}) is $35$ bits. Following an HCP, for the
data illustrated in Fig.\ref{corr_vanish}, this quantity becomes $29$ bits. It can be seen from
Fig. \ref{corr_vanish} that a large part of the phase information associated with $31p$ is lost.

When we apply a second HCP at a delay $T_2$ following the first HCP, it causes a redistribution of
the states into the $p$-states depopulated by the first HCP in a coherent manner and we can once
again measure the phase information in the previously missing state (see Fig.
\ref{corr_recovered}). The effect of the second HCP is to recover the information hidden by the
first HCP. The uncertainty in the phase of the $31p$ state is reduced by nearly a factor of three
from $32^\circ$ to $12^\circ$ ($i=4.9$ bits).

The timing of the second HCP affects not only the final phase and amplitude of the depopulated
state but also those of the other $p$-states. The second HCP will in general tend to depopulate
those other $p$-states as well. At the delay shown in Fig.\ref{corr_recovered}, this produces a
slight increase in the uncertainties of the phases of the other $p$-states (corresponding to less
efficient information retrieval); this counteracts the information increase seen in $31p$. The
total information content ($28$ bits) is nearly unchanged for this delay of the second HCP. The
improvement that we observe is in the increase in information recovered from the $31p$ state.

We have also performed a separate experiment to determine whether the population transferred into
the $31p$ state is due to $p$-state redistribution alone or if it is also the result of population
transfer from non-$p$ states as we expected. We isolate the effects of neighboring $\ell\neq 1$
states on the recovery of correlations by exciting a two-state wave packet where the two excited
states are energetically distant, in our case, $27p$ and $32p$. Neither state has any $p$-state
neighbors to which it is coupled by the weak HCP.

The effect of the HCP on the lower energy $27p$ state is minimal, while the same HCP causes
significant transfer of $32p$ state amplitude into the neighboring $\ell\neq1$ states, namely
$31d$, $33s$, and $32s$. The final populations of the non-$p$ states are small, relative to the
$32p$ population. Note that what we measure as $32p$ population also contains $31d$ population;
the states are nearly degenerate, and we do not resolve them using ramped field ionization.

In the two-state experiment, none of the population was transferred by the first HCP into other
$p$ states. Any population transferred back into $32p$  as the result of a second HCP must
originate from a non-$p$ state. Meanwhile, the effect of the HCP on the lower energy eigenstate
($27p$) is minimal. The correlation thus becomes a measure of the actual phase shift in $32p$.

\begin{figure}[t]
\hspace{1mm}
\includegraphics[width=1.08\columnwidth]{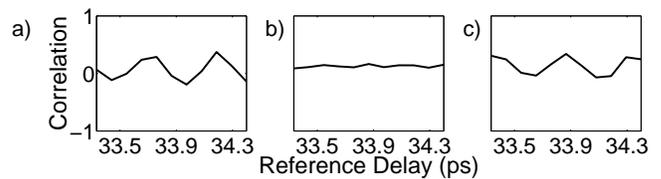}
\caption{\label{twostate_1} Two-state wave packet correlations.} a) In the absence of an HCP, the
$27p$-$32p$ correlation is strong; the phase relationship is well-defined and easily read. b)
$T_1=7\,\mathrm{ps}$. The operation of the HCP moves phase information into the $31d$ and $32s$
states, and that phase information becomes unreadable. c) $T_2=14.2\,\mathrm{ps}$. Application of
a second HCP at a later delay recovers phase information in the $32p$ state.
\end{figure}
\begin{figure}[t]
\includegraphics[width=\columnwidth]{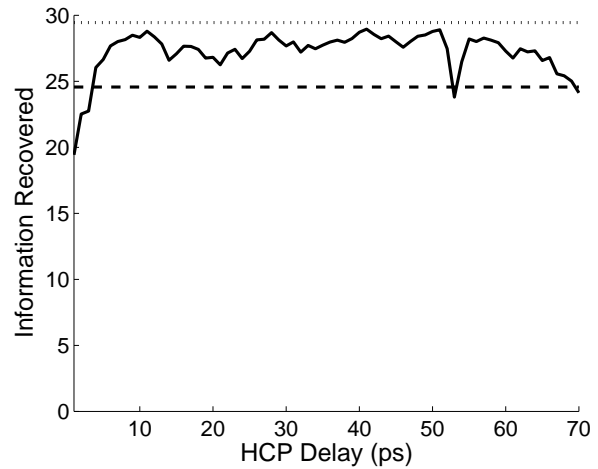}
\caption{\label{info_vs_delay2} Information recovered following second HCP.} In the absence of any
HCP, the information recovered is at the level indicated by the dotted line. An HCP arrives at a
fixed delay $T_1=4.1\,\mathrm{ps}$ after wave packet excitation, and the total phase information
retrieved following this HCP kick is indicated by the dashed horizontal line. Following the first
HCP, a second HCP kicks the wave packet at various delays, $T_2$. The information recovered
following the second HCP (solid line) is shown as a function of the time difference between the
first and second kicks.
\end{figure}

Since only two $p$-states are initially excited, there is only a single correlation to be
measured. In the absence of any HCP, the phase difference is well-defined (see
Fig.\ref{twostate_1}a). Upon application of a single HCP, the correlation amplitude between $32p$
and $27p$ vanishes (Fig.\ref{twostate_1}b). The phase information is hidden because of
depopulation of the $32p$ state in the presence of noise, as has been previously
established\cite{murrayPRA05}. When an HCP kicks the wave packet a second time, the coherence
between the $32p$ and $27p$ states is again observed (Fig.\ref{twostate_1}c), as a result of
transfer of coherent population from non-$p$ states into the partially depopulated $32p$ state.
This experiment proves that we can coherently transfer population from states that are accessible
to our measurement to states that are inaccessible and then retrieve it at a later time.

We consider the robustness of the retrieved information as a
function of the delay between the two HCP's. The first HCP is
applied at $T_1=4.1\ \mathrm{ps}$ such that it significantly
depopulates one of the states in our wave packet. In
Fig.\ref{info_vs_delay2}, the information capacity of the wave
packet after a single HCP is represented by the dashed line. With
the delay of the first HCP fixed, a second HCP arrives at various
delays and the total information content following the two HCPs' is
plotted as a function of this delay (solid line). It is seen that at
nearly all delays the second HCP can reliably recover the
information hidden by the first HCP.

The fundamental idea involved in the information hiding and retrieval scheme is that of
transferring state amplitudes into different subspaces (those with $\ell\neq1$), and transferring
them back into a particular manifold which we are able to measure experimentally. As a further
goal, we seek to learn about the content of states in a larger Hilbert space than we can measure
directly. In the present work, both the excitation of the larger Hilbert space and its probe were
half-cycle pulse operators. More generally, we find that we have a system which spans a large
state space, of which we are only able to directly measure a small fraction. Such a scenario need
not be limited to Rydberg atoms, but might include molecules and other multi-level systems.

This work has been supported by the National Science Foundation under grant no.9987916 and the
Army Research Office under grant no. DAAD 19-00-1-0373.

\bibliography{Infohiding-paper}

\end{document}